\documentclass[12pt]{article}
\usepackage{graphicx}
\usepackage{amsmath}

\title{ A New Approach for Reconstructing SUSY Particle Masses with a few $\rm fb^{-1}$ at the LHC}

\author{\textbf{Rashid M. Djilkibaev$^{1}$ \thanks{Permanent address:
     Institute for Nuclear Research, 60-th Oct. pr. 7a,
     Moscow 117312, Russia}\ ,
     Rostislav V. Konoplich$^{1,2}$}\\
   \normalsize$^{1}$Department of Physics, New York University,
   New York, NY 10003\\
   \normalsize$^{2}$Manhattan College, Riverdale, New York, NY, 10471}

\begin{document}

\maketitle

\begin{abstract}
We describe a  new cascade mass reconstruction approach  to allow 
reconstruction of  SUSY particle masses in long 
cascades (five or more particles) at the LHC 
with integrated luminosity as low as a few $\rm fb^{-1}$. 
This approach is based on a consecutive use of the endpoint method,
an event filter and a combinatorial mass reconstruction method. 
The endpoint method gives a preliminary estimate of light
sparticle masses.
An event filter combining the maximum likelihood distributions for all events in 
the data sample allows suppression of backgrounds and gives a 
preliminary estimate of heavy sparticle masses.
Finally, SUSY particle masses are reconstructed by a search  
for a maximum of a combined likelihood function constructed for each possible
combination of five events in the data sample. 

SUSY data sample sets for the SU3 model point containing 80k events 
each were generated, 
corresponding to an integrated luminosity of 4.2 $\rm fb^{-1}$.
These  events were passed through the AcerDET detector
simulator, which parametrized the response of a detector.
To demonstrate the stability and precision of the approach five different 80k event data sets 
were considered. 
Masses were reconstructed with  a precision of about $10\%$ for heavy sparticles  
and  $10-20\%$  for light sparticles.
\end{abstract}

\newpage

\section*{I. Introduction} 

   Supersymmetry \cite{susy} is a very popular extension of the Standard Model. 
By introducing an equal number of fermion and boson degrees of freedom
SUSY leads to a cancellation of quadratic divergences in the Higgs sector and
provides an attractive solution to the hierarchy problem \cite{hierar}. If 
supersymmetry exists at an energy scale of ~1 TeV, the production cross sections
of SUSY particles can be significant and effects of SUSY particles 
should be observed at the LHC, which should reach an integrated luminosity of
300 $\rm fb^{-1}$ in about 5 years \cite{atlas}.
   In many supersymmetric models, in order to avoid undesirable weak scale 
proton decay, R-parity conservation is assumed. This leads to pair 
production of SUSY particles and  therefore the lightest SUSY particle (LSP) is
stable. As a result, SUSY events at an accelerator will give two
decay chains, each containing one LSP and Standard Model particles in the
final state. The reconstruction of a SUSY event is complicated because of 
escaping LSPs and many complex and competing decay modes.
   In this paper we will consider mass reconstruction of five SUSY particles 
in the cascade decay

\begin{equation}
\tilde{g} \to \tilde{b} b_{2} \to \tilde{\chi}_{2}^{0} b_{1} b_{2} \to \tilde{l}_{R} l_{2} b_{1} b_{2} \to \tilde{\chi}_{1}^{0} l_{1} l_{2} b_{1} b_{2} 
\label{chain} 
\end{equation}

\begin{figure}[h]
  \centering
  \includegraphics[width=0.7\textwidth]{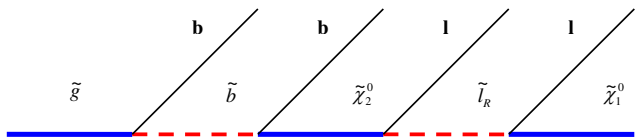} 
  \caption[Short caption.]{A gluino cascade decay chain.}
\label{fig:chain}
\end{figure}

The gluino decay chain (\ref{chain}) is shown in Fig.(\ref{fig:chain})

At present there are two different approaches to SUSY mass reconstruction.
The standard technique for analysis is ``the endpoint
method" which has been widely studied 
\cite{atlas} - \cite{lester} for high integrated
luminosity of about $100-300 ~\rm fb^{-1}$ at the LHC. This method looks 
for kinematic endpoints. SUSY particle masses are reconstructed by 
minimizing the $\chi^2$ function of the difference between observed and 
theoretical endpoints. 

The second method of SUSY particle mass reconstruction  
is the so called ``mass relation method" \cite{tovey},\cite{nojiri} based on the mass relation 
equation (see Appendix A) which relates SUSY particle masses and 
measured momenta of detected particles.
The ``mass relation equation" is
obtained as a solution  of a system of four-momentum
constraints for each vertex in the decay chain (\ref{chain}).
For example for the gluino decay vertex it follows that
$m_{\tilde{g} }^2 = (p_{\tilde{\chi}_{1}^{0}} + k_{l_{1}} + k_{l_{2}} + p_{b_{1}} + p_{b_{2}})^{2}
$
which give the relation between four components of LSP four-momentum in terms 
of decaying particle mass, LSP mass and the four-momenta of detectable particles. 
Similar relations can be obtained for each vertex in the process (\ref{chain}). 
There are four vertices in the cascade (\ref{chain}) so one gets four  kinematic 
equations that 
can be solved for four components of LSP four-momentum in terms of SUSY masses 
and momenta of detectable particles. By substituting these components into 
the on-shell mass condition for LSP
$m_{\tilde \chi_{1}^{0}}^{2} = p_{\tilde \chi_{1}^{0}}^{2}$
the ``mass relation equation" including all SUSY masses ($\vec m$)  and 
detectable particle momenta ($\vec p$) in the process (\ref{chain})
\begin{equation}
f( \vec m, \vec p) = 0 
\label{constr} 
\end{equation}
is found (an explicit form of this equation is given in Appendix A). 
This equation is valid for any SUSY event including the cascade 
(\ref{chain}). 
For one event this equation is underconstrained, it contains five unknown masses.

Authors of ~\cite{nojiri} developed two approaches to the problem 
of mass search at high luminosity of about $300 ~\rm fb^{-1}$.
Both approaches assumed that
three light SUSY particle masses  ($\tilde{\chi}_{2}^{0}$, $\tilde l_R$, 
and $\tilde{\chi}_{1}^{0}$) are known.
In this case, the mass relation equation contains only two unknown parameters,
the masses of the heaviest sparticles.
In the first ``event pair analysis" approach it was assumed that 
SUSY particles are the same from event to event.
Gluino and bottom squark masses are reconstructed in this case
by solving the system of two mass relation equations.
The second approach  is based on the search for 
the maximum of the combined likelihood function
for all events which is constructed by an individual likelihood
function for each event taking into account the mass relation 
equation in two dimensional $(m_{\tilde g},m_{\tilde b})$ mass
space. The search for the maximum of the combined 
likelihood function allows  reconstruction of  gluino and bottom squark 
masses at high integrated luminosity.

We note that the reconstruction at low 
integrated luminosity (with a few $\rm fb^{-1}$)
does not work for heavy sparticle masses for either  
``the endpoint method" and ``the mass relation method" in their original
forms. 

``The endpoint method" uses only 
events near the end point;  therefore, high statistics is required.
Also, in order to reconstruct the gluino mass a non-relativistic approximation
is used for the LSP near the end point. This approximation is not
justified for situations with $\tilde l_R$ close in mass either to
$\tilde{\chi}_{2}^{0}$ or $\tilde{\chi}_{1}^{0}$. Note that 
in \cite{gjelsten} an approach was proposed to obtain the gluino mass
without a non-relativistic approximation.

``Pair analysis" of ``the mass relation method" assumes that masses 
of SUSY particles are the same from event to event while  in reality heavy
particle masses follow a Breit-Wigner distribution. Therefore, an attempt
to extend this approach to five masses often leads to  
an inconsistent system of equations or wrong solutions.
An attempt to extend the original combined  
likelihood function approach  \cite{nojiri}  based on a grid search in five 
dimensions 
leads to unreasonably intensive computing calculations.

The goal of the present article is to develop an approach allowing 
reconstruction all of five masses of the cascade (\ref{chain}) at low 
integrated luminosity of a few $\rm fb^{-1}$. 
This luminosity can be reached at the early stage of the LHC in comparison with
a projected integrated luminosity of $300 ~\rm fb^{-1}$ in five years.
This approach is based on a consecutive use of the endpoint method,
an event filter and a combinatorial mass reconstruction method. 
It was found that reliable results are obtained only with a good starting 
point for the final fit and by rejecting a large fraction of the background.
The first two methods are used to define five SUSY particle mass ranges 
for the final combinatorial mass reconstruction method in which all five masses
are simultaneously fit and allowed to vary from event to event.
The endpoint method therefore is used to get a
preliminary estimation of 
the light SUSY particle ($\tilde{\chi}_{2}^{0}, \tilde l_R, \tilde{\chi}_{1}^{0} $) 
masses and corresponding errors.
These masses are then used with the mass relation equation constraint
to construct the maximum likelihood distribution in the two heaviest-sparticle mass-plane
for each individual event. 
An event filter combining the maximum likelihood distributions for all events in 
the data sample allows determination  of the range of heavy ($ \tilde{g}, \tilde{b}$)
masses  and significant suppression of the background.
Finally SUSY particle masses are reconstructed by a search  
for a maximum of a combined likelihood function,  
which depends on all five sparticle masses, 
constructed for each possible
combination of five events in the data sample. 
The approach is self-contained; no information on the nature of the
data set (SU3, SPS1a or other) is used.

Note that a possible way to improve ``the mass relation method''
was proposed in \cite{lester1} by considering all events
at the same time and fitting all five masses simultaneously.

\section*{II.Simulation }

We choose for this study the SU3 model point. 
This point has a significant production cross section for the 
chain (\ref{chain}): gluinos and squarks should be produced
abundantly at the LHC.
The bulk point SU3 
is the official benchmark point of the ATLAS collaboration 
and it is in agreement with
the recent precision WMAP data ~\cite{wmap}. 
This model point is described by the
set of mSUGRA \cite{msugra} parameters given in Table (\ref{tab:paramth}).

\begin{table} [h]
\begin{center}
 \begin{tabular}{|c|c|c|c|c|c|}
   \hline
   Point & $m_{0}$ & $m_{1/2}$ & $A_{0}$ & $tan\beta$ & $\mu$ \\
   \hline\hline
   SU3 & 100 GeV & 300 GeV & -300 GeV & 6 &  $>$ 0 \\
   \hline
 \end{tabular}
\caption{mSUGRA parameters for the SU3 point.}\label{tab:paramth}
\end{center}
\end{table} 

Branching ratios for the gluino decay chain (\ref{chain}) at the SU3 point are

\begin{equation}
\nonumber
\tilde{g} \stackrel{16.6 \%}{\longrightarrow} \tilde{b}_{1} \stackrel{24.1 \%}{\longrightarrow} \tilde{\chi}_{2}^{0} 
\stackrel{11.4 \%}{\longrightarrow} \tilde{l}_{R} \stackrel{100 \%}{\longrightarrow} \tilde{\chi}_{1}^{0} ~~\Rightarrow ~~0.46 \% 
\end{equation}

\begin{equation}
\nonumber
\tilde{g} \stackrel{9.2 \%}{\longrightarrow} \tilde{b}_{2} \stackrel{16.6 \%}{\longrightarrow} \tilde{\chi}_{2}^{0} 
\stackrel{11.4 \%}{\longrightarrow}  \tilde{l}_{R} \stackrel{100 \%}{\longrightarrow} \tilde{\chi}_{1}^{0} ~~\Rightarrow ~~0.18 \% 
\end{equation}

\noindent 

Assumed theoretical masses of SUSY particles in the cascade (\ref{chain}), the total branching ratio  
and a cross section
generated by ISAJET 7.74 are given in Table (\ref{tab:massth}).

\begin{table} [h]
\begin{center}
 \begin{tabular}{|c|c|c|c|c|c|c|c|c|}
   \hline
   Point & $m_{\tilde{g}}$ & $m_{\tilde{b_{1}}}$ & $m_{\tilde{b}_{2}}$ & $m_{\tilde{\chi}_{2}^{0}}$ & 
   $m_{\tilde{l}_{R}}$ & $m_{\tilde{\chi}_{1}^{0}}$ & BR& $\sigma [\rm pb]$ \\
   \hline\hline
   SU3 & 720.16 & 605.93 & 642.00 & 223.27 & 154.63 & 118.83 & 0.64$\%$ &19\\
   \hline
 \end{tabular}
\caption{Assumed  theoretical masses of sparticles, branching ratio BR and production cross section $\sigma $ 
at the SU3 point. 
Masses are given in GeV.}\label{tab:massth}
\end{center}
\end{table}  
Monte Carlo simulations of SUSY production at model points were performed by the HERWIG 6.510
event generator  ~\cite{herwig}. The produced events were passed through the AcerDET 
detector simulation ~\cite{atlfast}, which parametrized the response of a 
detector. Samples of 80k  SUSY events were used. 
This approximately  corresponds  to $4.2 ~\rm fb^{-1}$ of integrated luminosity
because the SUSY production cross section is 19 pb 
at the SU3 point.
Five different sets of 80k  SUSY events 
were considered to demonstrate the stability and precision of the mass reconstruction 
approach.

In order to isolate the chain (\ref{chain}) the following  
standard cuts were applied:

$\bullet$ two isolated opposite-sign same-flavor (OSSF) leptons (not tau leptons)
satisfying transverse momentum cuts $p_{T}(l^{\pm}) > 20~GeV$ and 
$p_{T}(l^{\mp}) >  10~GeV$

$\bullet$ two b-tagged jets with $p_{T} > 50~GeV$; 

$\bullet$ At least three jets, the hardest satisfying 
$p_{T1} > 150~GeV$, $p_{T2} > 100~GeV$, $p_{T3} > 50~GeV$;  

$\bullet$ $M_{eff} > 600~GeV$ and $E_{T}^{miss} > 0.2M_{eff}$, where
$E_{T}^{miss}$ is the missing transverse energy and $M_{eff}$ is the scalar
sum of the missing transverse energy and the transverse momenta of the four
hardest jets; 

$\bullet$ lepton invariant mass $50~GeV < M_{ll} < 105~GeV$.

Note that at the first stage of the reconstruction procedure to 
estimate light masses ($\tilde{\chi}_{2}^{0}$, $\tilde l_R$, 
and $\tilde{\chi}_{1}^{0}$), the chain
$\tilde{q_L} \to \tilde{\chi}_{2}^{0} q \to \tilde{l}_{R} l_{2} q \to \tilde{\chi}_{1}^{0} l_{1} l_{2} q$
was considered with the same cuts as mentioned above except the 
requirement of b-tagged jets and without a cut in lepton invariant mass. 
Lepton invariant mass $M_{ll}$ cuts were
determined from analysis of $ll$ distribution for the above $\tilde{q_L} $ chain 
which gives the upper edge 
with relatively high precision. $M_{ll}$ cuts reject about 1/4 of signal
events but improve the signal to background ratio.

It was shown in \cite{gj_note} that the QCD processes are cut down by the 
requirement of two leptons and of considerable missing $E_{T}$. The processes
involving Z and W are suppressed by the requirement of high hadronic activity
together with high missing $E_{T}$. The only Standard Model background 
surviving the hard cuts mentioned above is $t\bar{t}$ production, where
both W's decay leptonically into a $bbll$ state. The $t\bar{t}$ background
contributes about $20\%$ to the total number of events with $b\bar{b}l^+l^-$
in the final state after cuts.We found that this number can be reduced 
to about $2\%$ by applying an event filter procedure described in chapter IV.
This contribution is negligible in comparison with more than $100\%$ 
contribution due to SUSY bacground and we do not include 
$t\bar{t}$ background in the foolowing analysis.

Table (\ref{tab:sigbcgsu3}) 
shows the number of signal events and SUSY
background events for the SU3 model point after cuts were applied to 
the five sets of 80k SUSY events that corresponds to an integrated luminosity
of 4.2$\rm fb^{-1}$. 
A classification of events as signal and SUSY background ones is based on truth
information.
The SUSY background to the process (\ref{chain})
is significant.
It follows from Table (\ref{tab:sigbcgsu3}) that for the SU3 point the number of SUSY 
background events is  a factor 2 greater than the number of signal events.

\begin{table} [htb!]
\begin{center}
 \begin{tabular}{|c|c|c|c|c|}
   \hline
   Set & Total & Signal & SUSY Backg. & Ratio  \\
   \hline\hline
   1 & 154 & 47 & 107 & 3.3\\
   2 & 131 & 48 & 83 & 2.7\\
   3 & 148 & 44 & 104 & 3.4\\
   4 & 157 & 50 & 107 & 3.1\\
   5 & 141 & 55 & 86 & 2.6\\
   \hline
   1-5 & 731 & 244 & 487 & 3.0 \\
   \hline
 \end{tabular}
\caption{The number of signal and SUSY background events after cuts applied to
80k SUSY events. Ratio = (Signal+Background)/Signal.
Row 1-5 shows the number of events for five sets combined.}\label{tab:sigbcgsu3}
\end{center}
\end{table}              

\begin{figure}[h]
  \centering{\hbox{
  \includegraphics[width=0.5\textwidth]{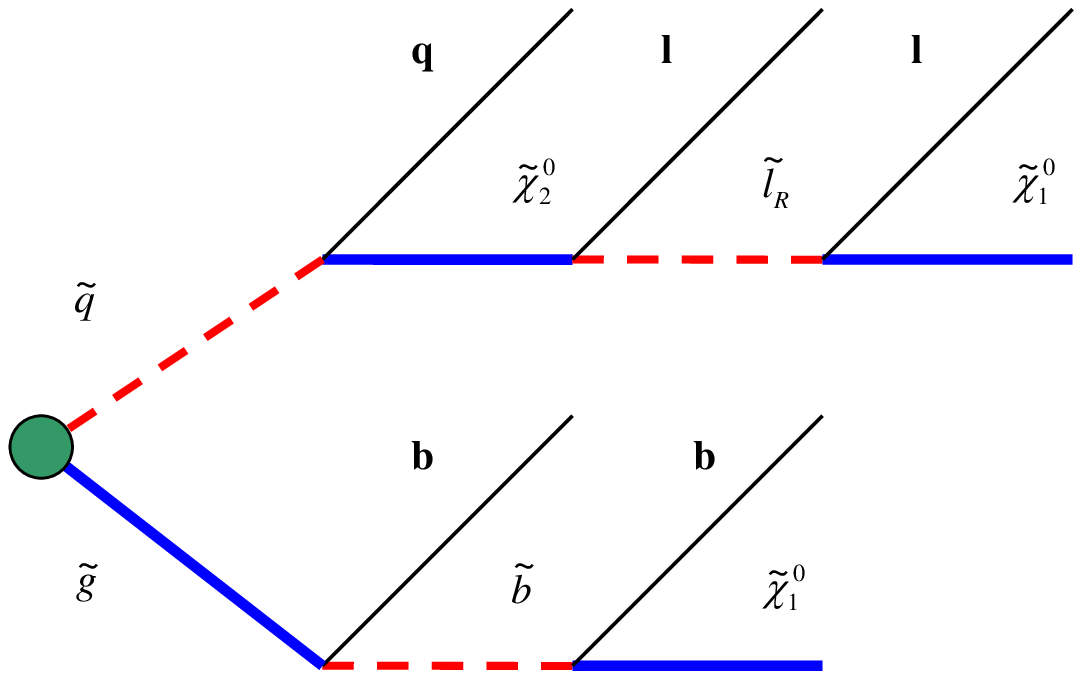} 
  \includegraphics[width=0.5\textwidth]{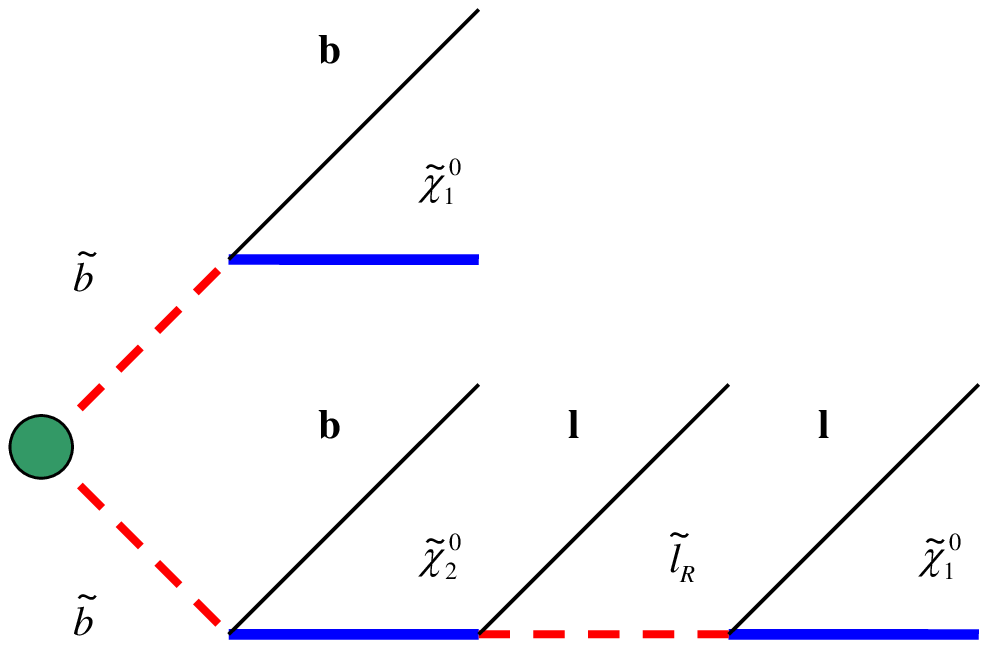} }}
  \caption[Short caption.]{Examples of SUSY background processes: neutralinos
appear in wrong vertices (left); gluino is missing (right).}
\label{fig_goodbcg}
\end{figure}

A better suppression of the SUSY background could be achieved from detail examination of 
background contributions.
Three different types of SUSY background can be considered:
a background containing $\tau$ leptons decaying into electrons or muons, 
a background with related processes in which  
$\tilde{\chi}_{2}^{0}$ decaying into $\tilde l_R$ appears in a wrong
vertex or if no gluino is produced  
as  shown in Fig.(\ref{fig_goodbcg}) and
other background. 
We estimate their relatives contributions at the SU3 point as $50\%, 25\%$ and 25 $\% $, respectively. 

A significant contribution to SUSY background comes from dominant 
modes of $\tilde g$, $\tilde b$, and $\tilde{\chi}_{2}^{0}$ decays. 
For the SU3 point $\tilde{\chi}_{2}^{0}$ decay into the pair
$\tau \tilde \tau_1$ is favored over the decay to $\tilde l_R$ due to the
relatively high value of $tan\beta$:
BR($ \tilde {\chi}_{2}^{0} \to {\tau}^{\pm} \tilde {\tau_1}^{\mp} \to {\tau}^{\pm} {\tau}^{\mp} \tilde {\chi}_{1}^{0} $)
equals $48.7\%$.
The presence of 
$\tau$ leptons  abundantly produced in 
dominant decay modes of SUSY particles 
would be a clear indication of SUSY background.
$\tau$-decays can create electrons or muons in the final state thus
faking the cascade (\ref{chain}).
The following decay sequences of gluino or bottom squark also
imitates the cascade (\ref{chain}) after $\tau$-decays into leptons:

$\tilde{g} \to \bar{\tilde{t_1}} t \to \bar{b} \tilde{\chi}_{1}^{-} bW^{+}  \to \bar{b} \tilde{\tau_1}^{-} \bar{\nu_{\tau}} 
bl^{+}\nu_l \to \bar{b} \tau^{-} \tilde{\chi}_{1}^{0} \bar{\nu_{\tau}} bl^{+}\nu_l$ ,

$\tilde{b_{1}} \to \tilde{\chi}_{1}^{-} t \to \tilde{\tau_1}^{-} \bar{\nu_{\tau}} bW^{+}  \to \tau^{-} \tilde{\chi}_{1}^{0} \bar{\nu_{\tau}} bl^{+}\nu_l$ .

The application of $\tau$ tagging could provide an additional 
suppression of combinatorial SUSY background by approximately a 
factor of 2.  
Note that in this work we did not use $\tau$ tagging for background
suppression.

For the second type of background (Figure \ref{fig_goodbcg}) 
only one b-quark appears in a wrong vertex which effectively leads to
smearing in momentum of the corresponding b-jet. As a result, for 
reasonably small smearing these processes, even though they are background, 
can give quite correct mass peak positions upon reconstruction.

\section*{III. Preliminary estimate of light masses by the endpoint method}

The endpoint method has been widely used to determine masses of 
SUSY particles ~\cite{endpoints} in particular in application to a decay chain

\begin{equation}
\tilde{q_L} \to \tilde{\chi}_{2}^{0} q \to \tilde{l}_{R} l_{near} q \to \tilde{\chi}_{1}^{0} l_{far} l_{near} q
\label{chainqL} 
\end{equation}

\noindent
which is a subprocess of the cascade (\ref{chain}) if one considers
$\tilde{q_L}$ instead of $\tilde{b}$. For leptons in the process 
(\ref{chainqL}) the following notations are used: 
$l_{near}$ stands for a lepton which is ``nearer'' to the quark 
while the lepton radiated by the slepton is called $l_{far}$
because it is ``further'' from the quark.

In the process (\ref{chainqL}) there are three detectable particles:
a quark and two leptons. Four invariant mass distributions can 
be formed from them: $\bf m_{ll}$, $\bf m_{qll}$, $\bf m_{ql_{near}}$, $\bf m_{ql_{far}}$.
The $\bf m_{ll}$ and $\bf m_{qll}$ distributions are observable, in particular,
the $\bf {ll}$ edge, $\bf {qll}$ edge and $\bf {qll}$ threshold can be measured. It is impossible to
observe $\bf m_{ql_{near}}$ and $\bf m_{ql_{far}}$ distributions because
the $\bf l_{near}$ and $\bf l_{far}$ assignment is ambiguous, however these
invariant masses can be combined to give $\bf {ql}$ low and $\bf {ql}$ 
high edges which are observable.

\subsection*{Endpoint extraction}

To extract endpoints from these distributions a fit was performed. The $\bf {ll}$
distribution was fitted in a standard way by a triangle form
convoluted with a Gaussian. 
Usually the four remaining distributions are fitted near the
edges by straight lines, but for 80k data samples this approach
does not work, due to low statistics. 
Instead predefined functions covering a wider range of the distributions (0 - 1 TeV)
have been used to get the endpoints. 
The $\bf {qll}$, $\bf {ql}$ low and $\bf {ql}$ high distributions
were fitted by a parabola and a straight line
in the range 0-1 TeV.
The $\bf {qll}$ threshold distribution was fitted by an exponential 
function and a straight line in the range 0-1 TeV. 

\begin{figure}[htb!]
  \centering
  \includegraphics[width=1.0\textwidth]{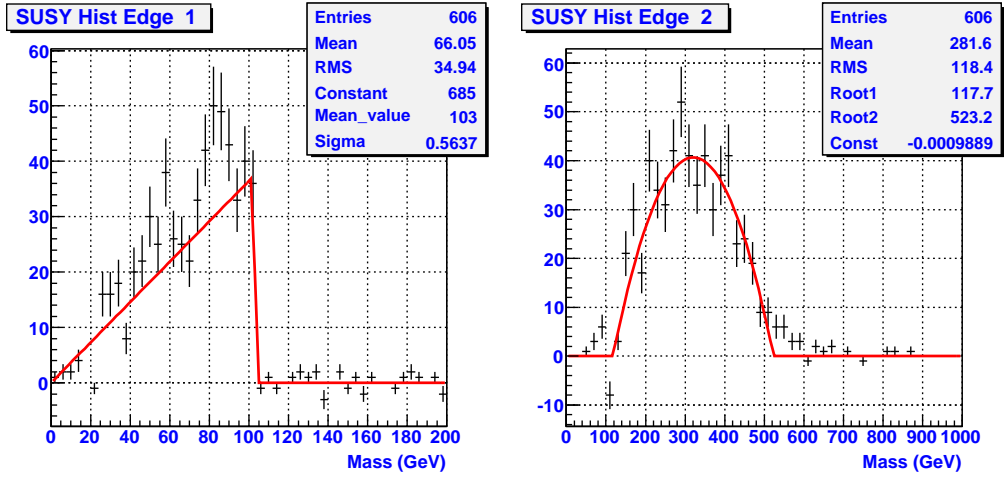} 
  \caption[Short caption.]{Invariant mass distributions for Set 1 at the SU3 point:  $\bf {ll}$ edge (left),
 $\bf {qll}$ edge (right)
}\label{fig_edge1}
\end{figure}

\begin{figure}[htb!]
  \centering
  \includegraphics[width=1.0\textwidth]{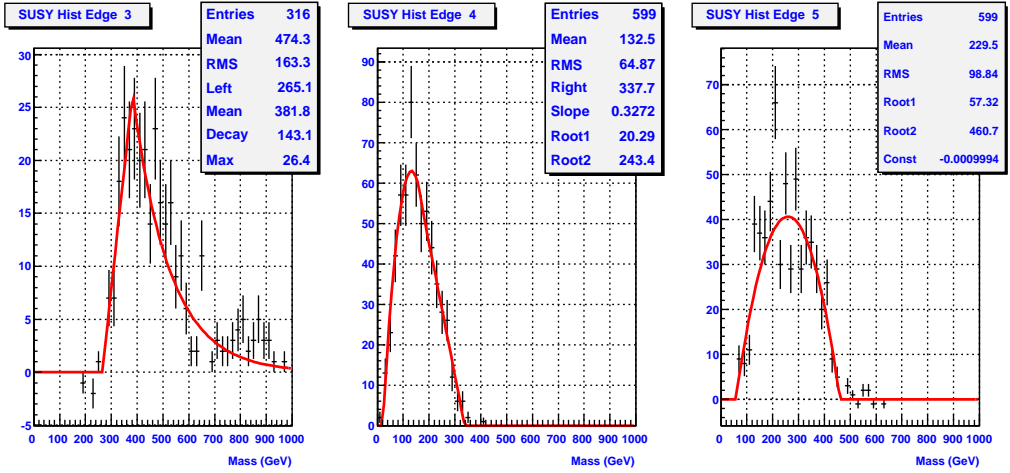} 
  \caption[Short caption.]{Invariant mass distributions for Set 1 at the SU3 point: 
  $\bf {qll}$ threshold (left), $\bf {ql}~$low edge (middle) and $\bf {ql}~$high edge (right)
}\label{fig_edge2}
\end{figure}

Figures (\ref{fig_edge1}) and (\ref{fig_edge2}) show five 
invariant mass 
distributions for Set 1 of 80k  events at the SU3 point
after applied the kinematic cuts and  OSOF subtraction. 
For Set 1
the endpoints and reconstruction errors found from fitting of the edges
are $\{$103$\pm$2, 523$\pm$6, 265$\pm$4, 338$\pm$6, 461$\pm$5$\}$ 
to be compared with heoretical endpoints $\{$103.1, 535.2,
263.8, 340.7, 456.0$\}$ for the SU3 point.

\subsection*{Mass region search}

Table (\ref{tab:edges}) gives the kinematic endpoints in dependence on
SUSY particle masses \cite{gjelsten0},\cite{gjelsten}.
The different cases listed for  $\bf {qll}$ and $\bf {ql}$
distributions are distinguished by mass ratios of sparticles.

\begin{table} [htb!]
\begin{center}
 \begin{tabular}{|p{3cm}|p{9.5cm}|}
   \hline
   Edge & Kinematic endpoints\\
   \hline\hline
   $\bf ll$ & $(\tilde \xi - \tilde l)(\tilde l - \tilde \chi)/\tilde l$\\
\hline
   $\bf qll ~threshold$ & $[(\tilde q + \tilde \xi)(\tilde \xi - \tilde l)
(\tilde l - \tilde \chi)+
2 \tilde l (\tilde q - \tilde \xi)(\tilde \xi - \tilde \chi)- 
(\tilde q - \tilde \xi)
\sqrt{(\tilde \xi + \tilde l)^2(\tilde l + \tilde \chi)^2 -
16 \tilde \xi \tilde \chi \tilde l^2}~]/(4 \tilde l \tilde \xi)$ \\
\hline
   $\bf {qll}$ & \\
 $case~ (1)$ & $(\tilde q - \tilde \xi)(\tilde \xi - \tilde \chi)/ \tilde \xi  
~ , ~~~~~~~~~~~~~ \tilde q/ \tilde \xi > \tilde \xi/ \tilde \chi$ \\
   $case~ (2)$ & $(\tilde q \tilde l - \tilde \xi \tilde \chi)(\tilde \xi - \tilde l)/(\tilde \xi \tilde l)
~ , ~~~~~~~~ \tilde \xi/ \tilde l > (\tilde l/\tilde \chi)(\tilde q/\tilde \xi)$ \\

   $case~ (3)$ & $(\tilde q - \tilde l)(\tilde l - \tilde \chi)/ \tilde l  
~ , ~~~~~~~~~~~~~~~ \tilde l/ \tilde \chi > \tilde q/ \tilde l$ \\
   $case~ (4)$ & $(\sqrt{\tilde q} - \sqrt{\tilde \chi})^2  
~ , ~~~~~~~~~~~~~~~~~~~ otherwise$ \\
\hline
   $\bf {ql}$~ low, ~$\bf {ql}$~ high & \\
$case~ (1)$ & $(m_{ql_{near}}^{max})^2 ~,~ (m_{ql_{far}}^{max})^2 
~,~~~~~~~ 2\tilde l/ \tilde \chi > \tilde \xi/ \tilde \chi + 1$\\
   $case~ (2)$ & $(m_{ql_{bound}}^{max})^2 ~,~ (m_{ql_{far}}^{max})^2 
~,~~~~~~~~~ \tilde \xi/ \tilde \chi + 1 > 2\tilde l/ \tilde \chi > 2\sqrt{\tilde \xi/ \tilde \chi} $\\
   $case~ (3)$ & $(m_{ql_{bound}}^{max})^2 ~,~ (m_{ql_{near}}^{max})^2 
~,~~~~~  2\sqrt{\tilde \xi/ \tilde \chi} > 2\tilde l/ \tilde \chi $\\
   \hline
 \end{tabular}
\caption{The kinematic endpoints for the process (\ref{chainqL}). The following notations are used
for masses
$\tilde \chi = m_{\tilde \chi_{1}^{0}}^2$~, $\tilde l = m_{\tilde l_{R}}^2$~, 
$\tilde \xi = m_{\tilde \chi_{2}^{0}}^2$~, $\tilde q = m_{\tilde q}^2$ and endpoints 
$(m_{ql_{near}}^{max})^2=(\tilde q - \tilde \xi)(\tilde \xi - \tilde l)/\tilde \xi$,
$(m_{ql_{far}}^{max})^2=(\tilde q - \tilde \xi)(\tilde l - \tilde \chi)/\tilde l$,
$(m_{ql_{bound}}^{max})^2=(\tilde q - \tilde \xi)(\tilde l - \tilde \chi)/
(2\tilde l -\tilde \chi).$ For $\bf {qll}$ and $\bf {ql}$ edges endpoints are given
by different expressions in dependence on mass ratios. 
}\label{tab:edges}
\end{center}
\end{table} 
\noindent
While the $\bf {ll}$ edge and $\bf {qll}$ thresholds are given by unique expressions, 
the other endpoints have different formulas for different
mass ratios which are not known $a ~priori$: there are four expressions for the $\bf {qll}$ edge and
three expressions for $\bf {ql}$ low and $\bf {ql}$ high edges. 
Each overall combination corresponds to a unique mass region in SUSY
particle mass space $m_{\tilde{q_L}}$, $m_{\tilde{\chi}_{2}^{0}}$, 
$m_{\tilde{l}_{R}}$, $m_{\tilde{\chi}_{1}^{0}}$. These mass regions 
defined by mass ratios from Table (\ref{tab:edges})
are labeled by R(i,j), where i=1,2,3,4 and j=1,2,3 denote the 
corresponding mass ratios for $\bf {qll}$ and $\bf {ql}$ endpoints, respectively.
According to the 
analysis performed in \cite{gjelsten} only nine of these 12 $(4 \times 3)$
combinations are physical, the regions R(2,1), R(2,2), and R(3,3)
are not possible. Additionally, the regions R(2,3), R(3,1), and R(3,2)
are degenerate:
the invariant mass distributions are not linearly independent,
because the $\bf {qll}$ edge is the function of the $\bf {ll}$ and $\bf {ql}$ high edges.

A procedure was developed to determine from the data sample the SUSY particle mass-space 
region.
Once the SUSY mass region is defined, the endpoint expression is determined in
a unique way from Table \ref{tab:edges}. 
Expressions for endpoints are inverted to get masses in terms 
of endpoints. Four endpoints
are required to invert the formulas because there are four unknown
masses in the process (\ref{chainqL}). For formula inversion
we always use  the $\bf {ll}$ distribution because it gives the endpoint with
relatively high precision and we take any three endpoints of the remaining four.
For example, one can extract four sparticle masses by considering
either $\bf {ll,~qll}$, $\bf {ql}$ low, $\bf {ql}$ high combination of endpoints 
or $\bf {ll,~qll}$, $\bf {ql}$ low, $\bf {qll}$ threshold combination and so on.
Thus for each region R(1,1), R(1,2), R(1,3), R(4,1), R(4,2), R(4,3) we
get four different sets of mass formulas. For 
degenerate regions R(2,3), R(3,1), R(3,2)
we get only two sets of mass formulas because there are only four
independent endpoints. 
Note that inverse formulas are very sensitive to the endpoint
determination precision because some formulas contain
singularities in the range of 30 GeV  around actual 
endpoints. The use of the full shapes of the signal as it was proposed in
\cite{raklev},\cite{lester2} could lead to an improvement in endpoint
extraction.   

The SUSY mass region is accepted if for all sets of inverse formulas the hierarchy condition 
$m_{\tilde{q_L}} ~>~ m_{\tilde{\chi}_{2}^{0}} ~>~ m_{\tilde{l}_{R}} ~>~ m_{\tilde{\chi}_{1}^{0}} ~>~ 20~GeV$   
is satisfied and found masses correspond to the mass ratios of Table (\ref{tab:edges}). 
Variations of low mass bound on the lightest neutralino mass in 
the range 20-50 GeV does not affect the results. Note that the
limit from accelerator experiments on $m_{\tilde{\chi}_{1}^{0}}$
is $46~ GeV$ \cite{limit}.   

For all five SU3 data sets  the R(1,3) region in mass space was found,
which is the true region for the SU3 point.
Note that the mass region search is sensitive to fluctuations in
invariant mass distributions. An increase in integrated luminosity
should lead to more precise reconstruction of light SUSY particle masses. 

\subsection*{Preliminary estimation of light SUSY particle masses}

Once endpoints are found and mass regions are defined, sparticle masses
can be determined by a minimization of $\chi^2 $. Parameters $\vec {m}$
of a $\chi^2$ fit are the four sparticle masses 
$m_{\tilde{q_L}}$, $m_{\tilde{\chi}_{2}^{0}}$, $m_{\tilde{l}_{R}}$, $m_{\tilde{\chi}_{1}^{0}}$.
The $\chi^2$, as a function of the free parameters $\vec {m}$, is then given 
by

\begin{equation}
\chi^{2}(\vec {m}) = \sum_{i=1}^{5} \frac {(Q_{i}^{obs} - Q_{i}(\vec {m}))^2} {\sigma_{i}^2 }
\label{fit}
\end{equation}

\noindent
Expression $Q_{i}(\vec {m})$ is the endpoint as a function of sparticle masses for the corresponding 
SUSY mass region as given in Table (\ref{tab:edges}). 
$Q_{i}^{obs}$ are the  observable
endpoints and their errors $\sigma_i$
determined in the above procedure.

The results of the fit 
for preliminary mass estimates and their errors 
based on found 
endpoints, edge reconstruction errors and mass regions 
are summarized in Table (\ref{tab:massfitsu3}).

\begin{table} [htb!]
\begin{center}
 \begin{tabular}{|c|c|c|c|c|c|c|}
   \hline
   Point & Particle & Set 1 & Set 2 & Set 3 & Set 4 & Set 5  \\
   \hline\hline
   SU3 & $\tilde \chi_{2}^{0}$ & 201$\pm$33 & 245$\pm$25 & 206$\pm$36 & 252$\pm$46 & 199$\pm$34\\
   & $\tilde l_{R}$ & 130$\pm$33 & 179$\pm$27 & 133$\pm$34& 184$\pm$47& 128$\pm$27\\
   & $\tilde \chi_{1}^{0}$ & 96$\pm$29 & 140$\pm$26 & 104$\pm$30& 147$\pm$46& 93$\pm$28\\
   \hline
 \end{tabular}
\caption{Light masses preliminary determined from fitting of edges for 
the SU3 point.}\label{tab:massfitsu3}
\end{center}
\end{table}

\section*{IV. Event filter}

An event filter is required before the final fit to determine the heavy SUSY particle ($\tilde g$, $\tilde b$) 
mass range and suppress  background.
It is very important to reduce combinatorial background because  at the final stage 
of the reconstruction procedure
all possible five event combinations are considered. 
The  combinatorial background is defined as five-event combinations 
that include at least one background event.
The number of five-event combinations including only signal events is
given by $N_s = n_s!/(5!(n_s-5)!)$ where $n_s$ is the number of signal
events in a data sample. The total number of five-event combinations
including backgrounds is given by $N_{tot} = (n_s+n_b)!/(5!(n_s+n_b-5)!)$ 
where $n_b$ is the number of background events in a data sample. 
The contribution of combinatorial background
with respect to all possible five signal events 
combinations $(N_{tot}-N_s)/N_s$ in the case of high $n_s$ 
can be approximated by $((n_s+n_b)/n_s)^5$ 
which is of about 240 for the SU3 point,
as  follows from 
Table \ref{tab:sigbcgsu3}.
At the event filter stage we can reduce the 5-dimensional mass space for each event to a 2-dimensional one, 
supposing that three light SUSY particle masses 
($\tilde \chi_{2}^{0}$, 
$\tilde l_{R}$, $\tilde \chi_{1}^{0}$)
are fixed and taken as the preliminary masses found by the endpoint
method in the previous chapter.  

The event filter procedure is based on an approximate
likelihood function with the mass relation constraint for each event

\begin{equation}
L(m_{\tilde g}, m_{\tilde b}) = \prod_{i=1}^4 ~
exp\left[-\frac{(p_i^{event}-p_i^{meas})^2}{2\sigma_{i}^2} \right]
\label{LH1}
\end{equation}
where index i 
runs over 
observed particles (two b-quarks and two
leptons) and labels the measured absolute momenta $p_{i}^{meas}$,  the uncertainties $\sigma_i$ 
in their measurement,
and the  event true absolute momenta $p_{i}^{event}$.
The approximate likelihood function (\ref{LH1}) takes into account only uncertainties in
lepton and b-jet energy measurements.  
Note that positions of each of two
b-jets and of each of two leptons in the decay chain (\ref{chain}) are unknown. 
It is quite simple to resolve the b-jets assignment because usually 
(about $96\%$ of the time) the b-jet 
with higher $p_T$ originates from $\tilde b$-quark decay, since for 
SU3 point at rest system
$ p_{Tb_{1}} = (m_{\tilde b}^{2} - m_{\tilde \chi_{2}^{0}}^{2})/2m_{\tilde b} \gg
p_{Tb_{2}} = (m_{\tilde g}^{2} - m_{\tilde b}^{2})/2m_{\tilde g}$. 
Therefore for each event we assume that the b-jet 
with higher $p_T$ originates from $\tilde b$-quark decay.
Leptons with higher $p_T$ are produced in both vertices with comparable 
probabilities. In this work for simplicity  it is assumed 
that leptons with higher $p_T$ originate from $\tilde l_R$  decay.
We use the following parametrization for 
$\sigma_i$ in equation (\ref{LH1}): 
for b-jets ~~ $\sigma/E = 0.5/\sqrt {E(GeV)} ~\oplus~ 0.03$ and 
leptons 
$\sigma/E = 0.12/\sqrt {E(GeV)} ~\oplus~ 0.005.$

To find a maximum of the likelihood 
function (\ref{LH1}) with the constraint (\ref{constr})
is the same as 
to search for a minimum of $\chi^2= -2 \cdot \rm {log L}~$ function:

\begin{equation}
\chi^2(m_{\tilde g}, m_{\tilde b}) =  \sum_{i=1}^4 \frac{(p_i^{event}-p_i^{meas})^2}{\sigma_{i}^2} + \lambda f(\vec m, \vec p)~
\label{chi1}
\end{equation}
In Eq.(\ref{chi1}) the mass relation constraint $f(\vec m, \vec p)$ (\ref{constr}) is taking into account
by the Lagrange multiplier $\lambda$.
We construct an event likelihood distribution calculating minimum of the $\chi^2$ function 
(\ref{chi1}) for each of 10$^5$ randomly defined points
in the ($\tilde g$, $\tilde b$) mass plane,
assuming that they are distributed
uniformly in the range 
$0.4-1.4 TeV$ and $0.3-1.3TeV$, respectively. 
This effectively corresponds
to a two dimensional grid with spacing  of $3GeV \times 3GeV$. 
We add points ($m_{\tilde g}$, $m_{\tilde b}$) 
with a weight equal to $(\chi_{cut}^2-\chi^2)/2$
to an event likelihood distribution histogram if 
the minimization procedure of the function (\ref{chi1})  converges, 
$\chi^2 (m_{\tilde g}, m_{\tilde b}) < \chi_{cut}^2=10$ 
and $f < 10^{-4} GeV^2$. 
The minimization iteration procedure is considered to have converged if the  relative difference
in $\chi^2_{event}$ for two consecutive  
iteration steps is less than $0.5\%$ and the total number 
of iteration steps does not exceed 20.  
For  signal  events the event likelihood distribution  
has a maximum in the region of the ($\tilde g$, $\tilde b$) mass plane correlated with
the true masses of $\tilde g$ and $\tilde b$. 
Thus signal events should give a peak in the region of true masses. 
For background events there is no strong
correlation of maximum likelihood distribution with true ($\tilde g$, $\tilde b$) masses.   
We define the combined event likelihood function as 
\begin{equation}
\rm {log} ~L_{comb}(m_{\tilde g}, m_{\tilde b}) = \sum_{events}{\rm {log} ~L(m_{\tilde g}, m_{\tilde b})}  
\label{chicomb}
\end{equation}
where the sum includes all events in the data sample.

\begin{figure}[htb!]
  \centering
  \includegraphics[width=0.45\textwidth]{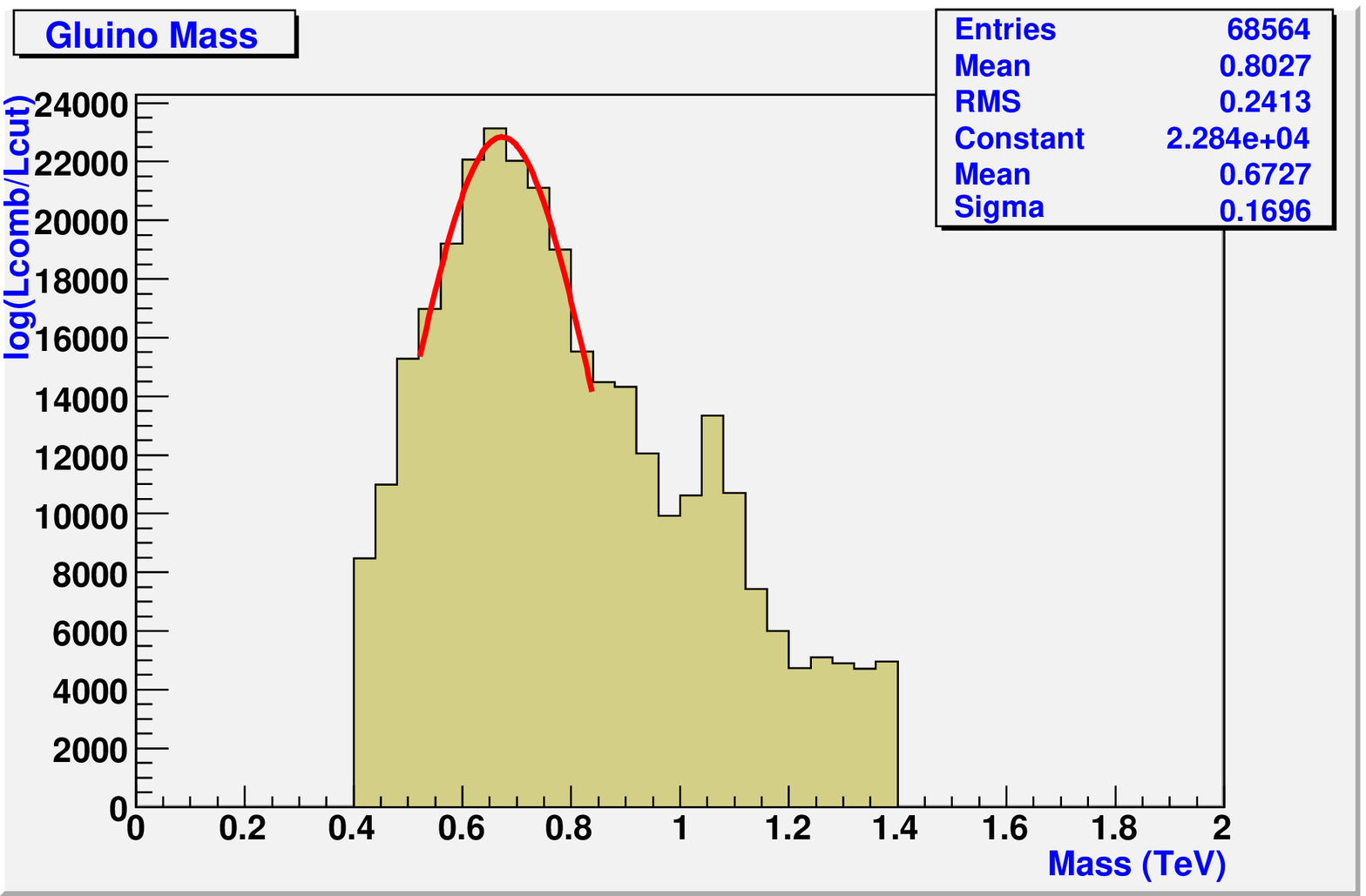}
  \hfill 
  \includegraphics[width=0.45\textwidth]{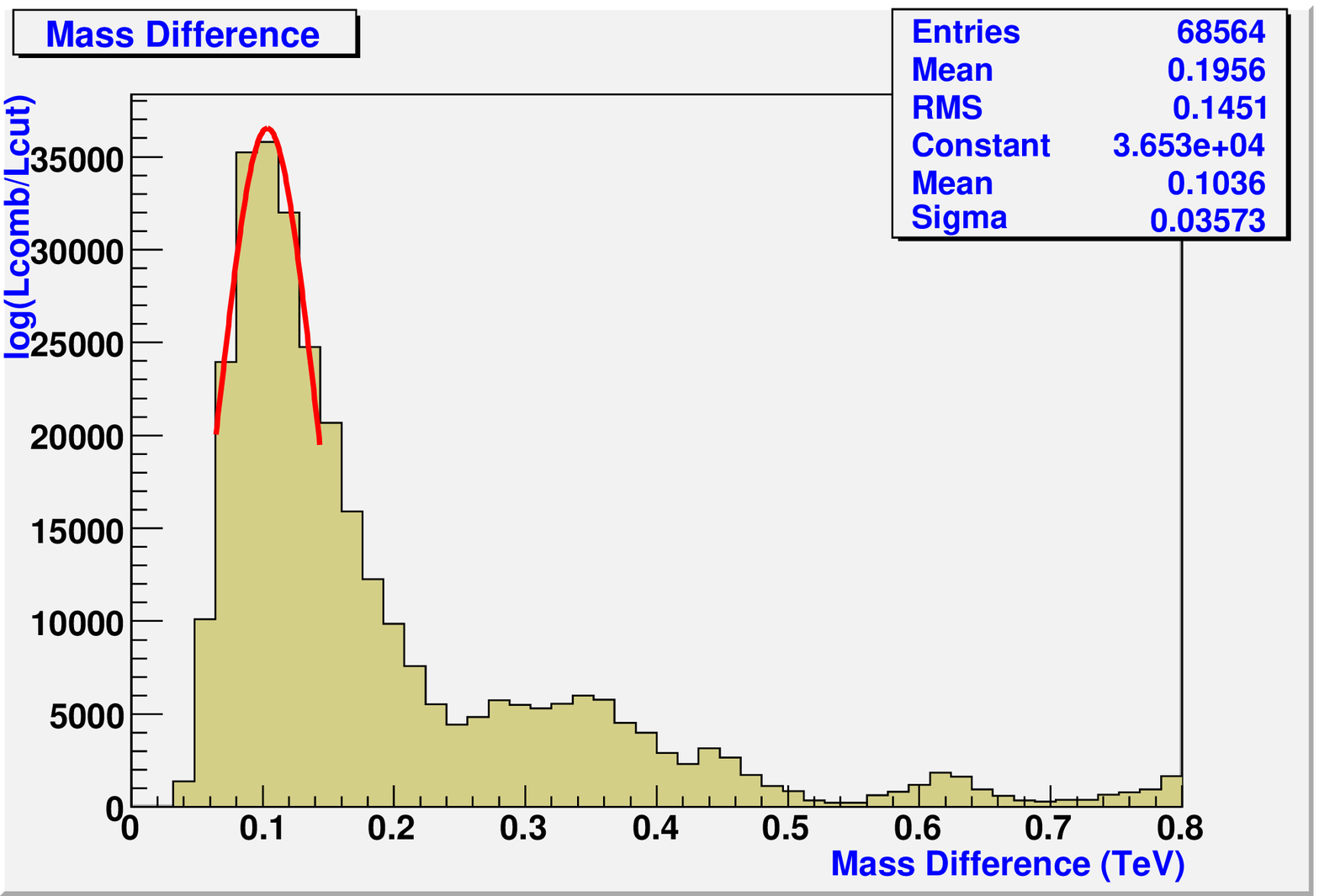} 
\caption{
 Combined event likelihood projection histograms for Set 1 at the SU3 point.
 }
\label{fig:gl_lhood}
\end{figure}

Figure (\ref{fig:gl_lhood}) shows two projection histograms of combined event likelihood distribution  
versus gluino mass $m_{\tilde g}$ and versus difference of gluino and 
sbottom  masses ($m_{\tilde g} - m_{\tilde b}$ for Set 1.
The variables ($m_{\tilde g}$, $m_{\tilde g} - m_{\tilde b}$) are used because
an event likelihood distribution is more sensitive
to these variables. 
Preliminary estimates for heavy sparticle masses and mass errors 
(the mean values  and standard deviations) 
obtained by 
fitting of these histograms by a Gaussian near the peak region are given 
in Table \ref{tab:msu3} for all five sets.

\begin{table} [htb!]
\begin{center}
 \begin{tabular}{|c|c|c|c|c|c|}
   \hline
   Particle & Set 1 & Set 2 & Set 3 & Set 4 & Set 5  \\
   \hline\hline
    $\tilde g$ & 673$\pm$170 & 720$\pm$124 & 683$\pm$158 & 729$\pm$211 & 649$\pm$110\\
    $\tilde b$ & 569$\pm$174 & 615$\pm$135 & 577$\pm$162& 621$\pm$214& 536$\pm$116\\
   \hline
 \end{tabular}
\caption{Preliminary heavy SUSY particle masses and mass errors estimated by event 
filter.}\label{tab:msu3}
\end{center}
\end{table} 

To suppress the background the previous procedure constructing 
an event likelihood distribution  
is repeated, but in the narrow  range in $m_{\tilde g}$ and $m_{\tilde b}$:
$m_{\tilde g} \pm \sigma$, $m_{\tilde b} \pm \sigma$ from Table \ref{tab:msu3}, assuming  
the uniform distributions.  
Requiring convergence and applying $\chi^2$ and the constraint f cuts as above results in
a likelihood distribution histogram  that has 
high efficiency for a signal event and low efficiency 
for background events.
Typically for a signal event the likelihood distribution histogram has 
of a few thousand entries (recall that 
the procedure is repeated $10^5$ times). 
In order to reject 
background and not suppress  signal, an entry threshold of 300 is
chosen. 
Thus if the event likelihood distribution histogram has 
greater than 300 entries the event is
considered a signal candidate and it is retained for the
consequent analysis.

After the application of the event filter the ratio of background
events to signal events is reduced approximately by a factor 2 as can be
seen from Table \ref{tab:efsu3} . 

\begin{table} [htb!]
\begin{center}
 \begin{tabular}{|c|c|c|c|c|c|}
   \hline
   Set &Total & Signal & SUSY &Ratio \\
    &Number & Events & Backg. & \\
   \hline\hline
   1  & 154/83& 47/39 & 107/44 & 3.3/2.1\\
   \hline
   2  & 131/83& 48/42 & 83/41 & 2.7/2.0\\
   \hline
   3  & 148/76& 44/35 & 104/41 & 3.4/2.2\\
   \hline
   4  & 157/110& 50/45 & 107/65 & 3.1/2.4\\
   \hline
   5  & 141/89& 55/45 & 86/44 & 2.6/2.0\\
   \hline\hline
   1-5  & 731/441& 244/206 & 487/235 & 3.0/2.1\\
   \hline
 \end{tabular}
\caption{The number of signal and background events before/after 
an application of event filter to 80k SUSY events. 
The last row shows the sum over  all five sets.}
\label{tab:efsu3}
\end{center}
\end{table} 

Contribution of the combinatorial
background which is given  approximately by ((Signal+Background)/Signal)$^5$
is therefore significantly suppressed. The suppression factor for the SU3 point
varies from 3 to 9.

\section*{V.  Combinatorial method for final mass reconstruction}

A combinatorial procedure is used for the final SUSY particle mass reconstruction.
It is applied only to the events that pass the event filter.
In order to explain this point let's neglect for a moment uncertainties
in detected particle momenta and the widths of the Breit-Wigner distribution. 
For a single event the mass relation constraint represents a four dimensional
surface in five-mass parameter space. Therefore, for two events the intersection
of two four dimensional surfaces is a three dimensional surface and correspondingly
for three events one gets a two dimensional surface of intersection, 
for four events the surface is one dimensional and at last for five events the 
intersection is just a point or a few points in five dimensional mass space
corresponding to SUSY particle masses. Thus in an ideal case five events 
would be enough \cite{nojiri} to reconstruct masses of SUSY particles.
Uncertainties in detected particle momenta and Breit-Wigner distribution
lead to smearing in the position of this point. 
There is a trade-off between the number of events in a combination and 
combinatorial background: more events restrict the number of degrees of 
freedom and better define SUSY masses, but 
combinatorial background increases. 
Therefore, at the final stage of mass reconstruction when the physical
background has already been reduced, we will consider all possible
five event combinations from the event sample.
Recall that in reality due to the Breit-Wigner distribution the masses of the gluino and bottom
squark vary from event to event.
A method described below takes into account that SUSY particle masses can vary from event to event.

SUSY particle masses are reconstructed by a search  
for a maximum of a combined likelihood function constructed for each possible
combination of five events in the data sample. 
For sparticle masses ($\vec m$) the combined likelihood function for the combination 
is defined as the product of the maximum likelihood functions for individual events. 
To find a maximum of the combined likelihood 
function for the combination is the same as 
to search for a minimum of the function 

\begin{equation}
\chi^2_{comb}(\vec m)=\sum_{i=1}^5{\rm {MIN} (\chi^2_{~event})_i} 
\label{chi20}
\end{equation}
where $\chi^2_{comb} = -2 \cdot \rm {log L_{comb}}$.
In Eq.(\ref{chi20}) $\rm {MIN} (\chi^2_{event})_i$ is a result of searching for a minimum of the 
$\chi^2_{event}$ function (\ref{chi2})
for an individual event for given $\vec m$ with the mass relation and ${\bf ll}$ edge constraints.
For each of the five events in the five event combination the $\rm {MIN} (\chi^2_{event})$ is fitted with
9 parameters (four  particle momenta and five SUSY masses) starting with
$\vec p^{~event} = \vec p^{~meas}$ and $\vec m^{event} = \vec m$.
The minimization iteration procedure is converged if a relative difference
in $\chi^2_{event}$ for two consecutive  
iteration steps is less than $0.5\%$ and the total number of 
iteration steps does not exceed 20.

The $\chi^2$ function for an event is defined by

\begin{equation}
\chi^2_{event} = \sum_{i=1}^4 \frac{(p_i^{event}-p_i^{meas})^2}{\sigma_{i}^2} 
+ \sum_{n=1}^5 \frac{(m_n^{event}-m_n)^2}{\sigma_{n}^2}  + \lambda_1 f + \lambda_2 f^{ll} ~.
\label{chi2}
\end{equation}
where the first term takes into account deviations of measured
momenta of b-jets and leptons from the true ones. The second term
takes into account that sparticles are varied from event to event and approximated by a Gaussian of width $\sigma_n$
instead a Breit-Wigner distribution.
In Eq.(\ref{chi2}) the mass relation and $\bf {ll}$ edge constraints are taking into account
by Lagrange multiplier $\lambda_{1}, \lambda_{2}$.
Standard deviations corresponding to the mass term are taking to be 15 GeV for the gluino,
5 GeV for bottom squark and 1 GeV for light masses. 
The first two numbers are comparable with
theoretical widths for heavy SUSY particles. The last number takes into account
the fact that light SUSY particles are quite narrow or stable. We note 
that the results of the mass reconstruction are not strongly sensitive to actual values   
of sparticle widths.

To get a starting point for the minimization of $\chi^2_{comb}$ 
for each five-event combination we calculate $\chi^2_{comb}$ 
for 3000 $\vec m$ values produced with a simple Monte Carlo.
The  set of the mass $\vec m$ with the smallest  $\chi^2_{comb}$  
is used as the starting point.
It is assumed
that heavy masses are distributed uniformly in the following range: 
mean value $\pm ~2\sigma$,
where the mean values and standard deviations 
are given in Table \ref{tab:msu3} as the results of event filter
application. For the three light masses the Gaussian distribution 
is assumed with mean values and standard deviations 
found by the endpoint technique as given in Table \ref{tab:massfitsu3}. 
The starting point is used by MINUIT code with the Simplex algorithm 
to search for the minimum of  $\chi^2_{comb}$ with five parameters $\vec m$.
The five-event combination is added to reconstruction mass histograms if 
the MINUIT minimization procedure  converges, $~\chi^2_{comb}  < \chi_{cut}^2=10$ 
and the sum of  the mass relation and $\bf {ll}$ edge constraint functions for five events is less than 1 $ GeV^2$. 
The CPU time required for the final minimization of the five-event combination is about 2 second for a 3GHz processor.
For a set of 100 events the required time for combinatorial mass reconstruction would be about 7 years.
In order to reduce the computational time, we divide a sample of 
80k  SUSY events into four subsets of 20k. 
Forming combinations only from events within each 20k subset greatly reduces combinatorics.
Because the total set is subdivided in four subsets 
the procedure is carried on for each subset and reconstruction mass histograms 
are merged.  

\begin{figure}[htb!]
  \centering
  \includegraphics[width=1.0\textwidth]{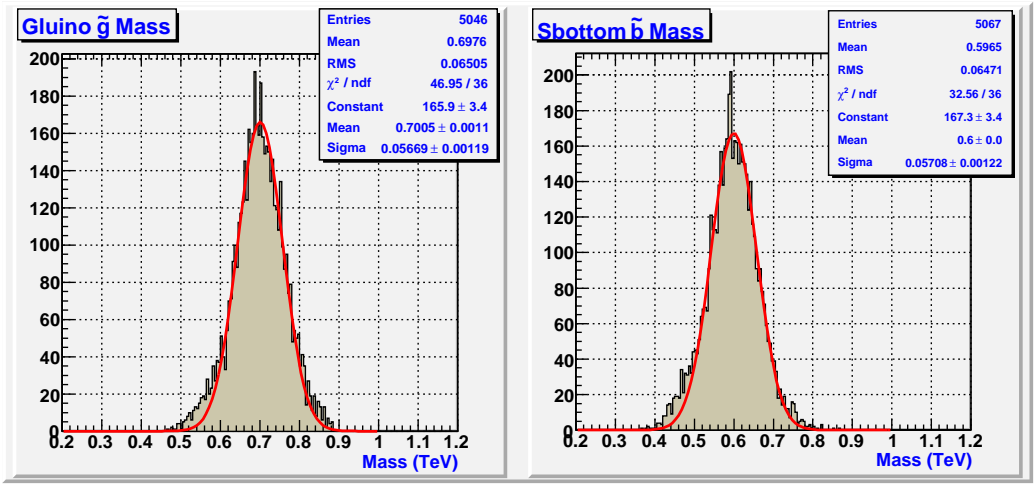} 
  \caption[Short caption.]{Reconstructed heavy SUSY particle mass distributions
  for Set 1, including background with integrated luminosity 4.2 $\rm fb^{-1}$:
 gluino $\tilde g$ (left),  sbottom $\tilde b$ (right). The line is a result of Gaussian fit.   
   }
\label{fig_mass1}
\end{figure}

\begin{figure}[htb!]
  \centering
  \includegraphics[width=1.0\textwidth]{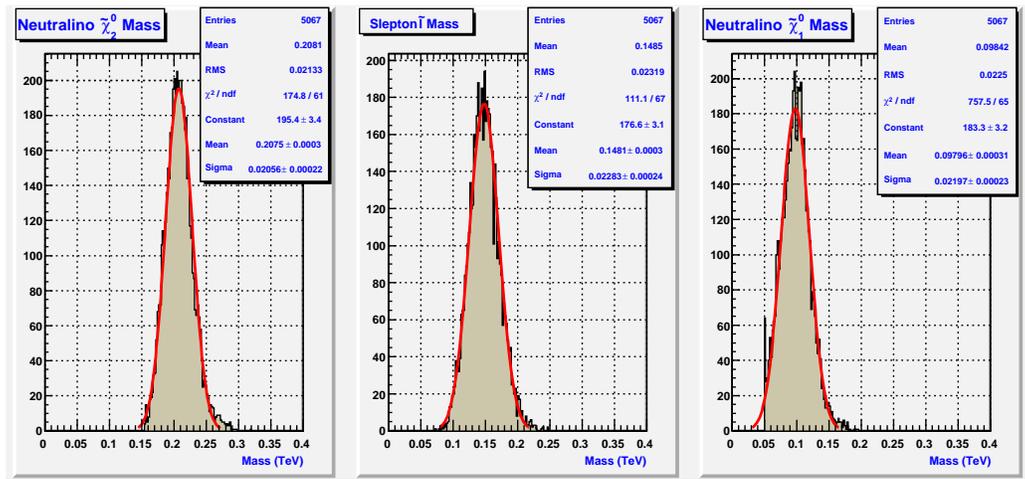} 
  \caption[Short caption.]{Reconstructed light SUSY particle mass distributions
 for Set 1, including background with integrated luminosity 4.2 $\rm fb^{-1}$:
 neutralino $\tilde \chi_{2}^{0}$ (left), slepton $\tilde l_{R}$ (middle), 
neutralino $\tilde \chi_{1}^{0}$ (right). The line is a result of Gaussian fit.  
   }
\label{fig_mass2}
\end{figure}

Reconstructed SUSY particle mass distributions and results of fitting these distributions by a Gaussian are
presented in Figures (\ref{fig_mass1} and \ref{fig_mass2}) for Set 1 at the SU3 point with integrated luminosity 4.2 $\rm fb^{-1}$.  
As can be seen in these Figures, the reconstructed mass distributions are well described by a Gaussian.
The mean values of the Gaussian fit are considered as the reconstructed sparticle masses.
As can been seen these masses are   close to the theoretical masses. 
It is assumed that mass reconstruction errors are given by the Gaussian fit standard deviations. 
For the light sparticles  
the final procedure improves mass reconstruction  errors by a factor of 2 but
the masses are changed only slightly in comparison with the preliminary estimate.

Final results  of this mass reconstruction approach 
are presented 
for five data sample sets of 80k events each 
in Table \ref{tab:reco}.

\begin{figure}[htb!]
  \centering
  \includegraphics[width=1.0\textwidth]{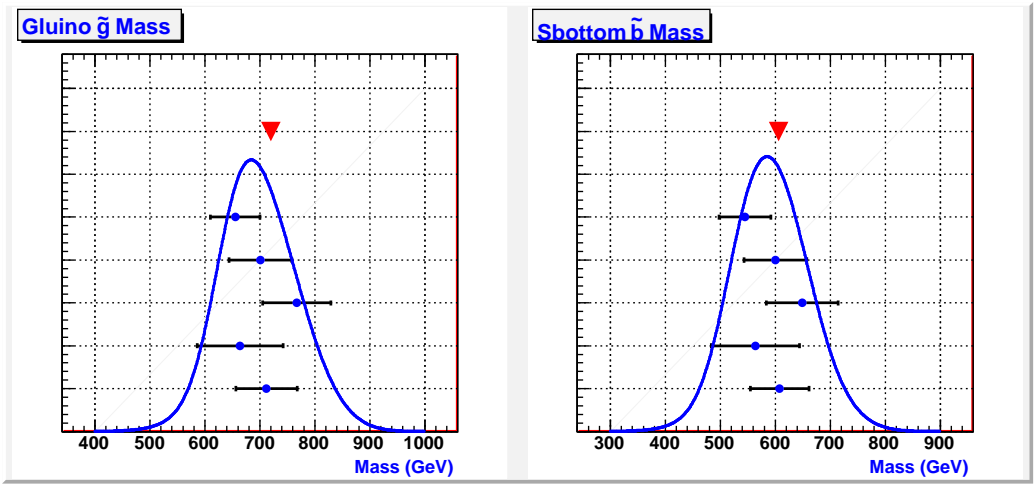} 
  \caption[Short caption.]{Ideograms of reconstructed heavy SUSY particle masses, 
including background for five data sample sets with integrated luminosity 4.2 $\rm fb^{-1}$:
 gluino $\tilde g$ (left),  sbottom $\tilde b$ (right). 
The triangle marker gives the position of theoretical mass.
Points with error bars correspond  to five data sample sets.

   }
\label{fig_ideo1}
\end{figure}

\begin{figure}[htb!]
  \centering
  \includegraphics[width=1.0\textwidth]{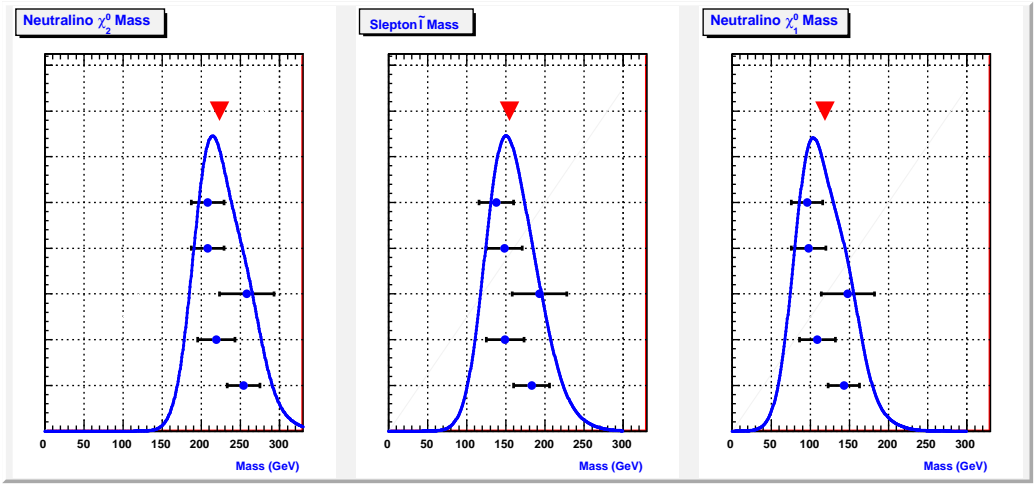} 
  \caption[Short caption.]{Ideogram of reconstructed light SUSY particle mass distributions,
including background for five data sample sets with integrated luminosity 4.2 $\rm fb^{-1}$:
 neutralino $\tilde \chi_{2}^{0}$ (left), slepton $\tilde l_{R}$ (middle), neutralino $\tilde \chi_{1}^{0}$ (right).  
Triangle marker gives the position of theoretical mass.
Points with error bars correspond  to five data sample sets. 
   }
\label{fig_ideo2}
\end{figure}

\begin{table} [htb!]
\begin{center}
 \begin{tabular}{|c|c|c|c|c|c|}
   \hline
   Set &$m_{\tilde{g}}$ & $m_{\tilde{b}}$ & $m_{\tilde{\chi}_{2}^{0}}$ & $m_{\tilde{l}_{R}}$ & $m_{\tilde{\chi}_{1}^{0}}$\\
   \hline
   1&701$\pm$57 &600$\pm$57 &208$\pm$21 &148$\pm$23 &98$\pm$22 \\
   2&712$\pm$55 &608$\pm$53 &254$\pm$21 &183$\pm$23 &143$\pm$20 \\
   3&664$\pm$78 &564$\pm$80 &219$\pm$24 &149$\pm$24 &109$\pm$23 \\
   4&767$\pm$62 &649$\pm$65 &258$\pm$35 &193$\pm$35 &148$\pm$34 \\
   5&655$\pm$45 &545$\pm$47 &208$\pm$21 &138$\pm$22 &96$\pm$20 \\
   \hline
 \end{tabular}
\caption{
Reconstructed SUSY particle masses and reconstruction errors for five data sample sets of 80k events each.
}
\label{tab:reco}
\end{center}
\end{table} 
In order to illustrate a spread in reconstructed masses
the results of Table \ref{tab:reco} are also shown in a form of ideograms \cite{graph}
in Figures (\ref{fig_ideo1} and  \ref{fig_ideo2})
for the five data sample sets. 
Each reconstructed mass in an  ideogram is represented by a Gaussian with a central value $m_i$, error $\sigma_i$ and
area proportional to 1/$\sigma_i$. The solid curve is a sum of these Gaussians.

The Gaussian form of ideograms and relatively small shift of peak positions
with respect to theoretical masses demonstrate the self-consistency of the mass 
reconstruction approach.

\section*{IX. Conclusion}

We have developed an approach that allows extracting all 
SUSY particle masses for long 
cascades at the LHC 
with an integrated luminosity of a few $\rm fb^{-1}$. 
This luminosity can be reached at the early stage of the LHC in comparison with
a projected integrated luminosity of $300 ~\rm fb^{-1}$ in five years.
This approach is based on a consecutive use of the endpoint method, 
an event filter and a combinatorial mass reconstruction method. 

The endpoint method allows  
preliminary estimate of light SUSY particle ($ \tilde \chi_{2}^{0}, \tilde l_{R}, \tilde \chi_{1}^{0}$) masses 
in a model-independent way. 

The light sparticle  masses are used 
to construct the maximum likelihood distribution in the two heaviest-sparticle mass-plane
for each individual event taking into account the mass relation equation
constraint. 
The event filter combining the maximum likelihood distributions for all events in 
the data sample allows a determination of the range of heavy sparticle
masses and significant suppression of background. This is
a new technique for background suppression. 
Note that in this
work we did not use $\tau$ tagging for background
suppression.
The application of $\tau$ tagging could provide an additional 
suppression of combinatorial SUSY background by approximately a 
factor of 2.  

Finally, SUSY particle 
$(\tilde g, \tilde b,
\tilde{\chi}_{2}^{0}, \tilde l_R, \tilde{\chi}_{1}^{0})$ masses
are extracted by a search  
for a maximum of a combined likelihood function constructed for each possible
combination of five events in the data sample satisfying the filter. Mass peaks 
for five sparticles are
clearly reconstructed because the remaining background does not exhibit peaks
corresponding to the signal region. 
With more events  the same technique could be used
to reconstruct, for
example, masses of the  $\tilde \tau_1$ and  $\tilde t_1$ in
the following cascade decays: $\tilde{g} \to \tilde{b} 
\to \tilde{\chi}_{2}^{0} \to \tilde{\tau}_{1} \to \tilde{\chi}_{1}^{0} $ and
$\tilde{g} \to \tilde{t_1} \to \tilde{\chi}_{2}^{0} \to \tilde{l}_{R} 
\to \tilde{\chi}_{1}^{0}$.  

SUSY data sample sets for this work 
were generated for the benchmark SU3 point
of mSUGRA scenario.
A detector response was parametrized by the AcerDET detector
simulator. 
The stability and precision of the approach was demonstrated by considering 
five different 80k event data sets for each model point. 
Masses were reconstructed with a precision of about $10\%$ for heavy sparticles  
and  $10\%, 15\%$ and $20\%$ for light $\tilde{\chi}_{2}^{0}, \tilde l_R$ and $\tilde{\chi}_{1}^{0}$ 
sparticles, respectively. 
The precision of mass reconstruction with this technique 
should be improved with an increase in integrated luminosity.

We also applied the cascade mass reconstruction approach to
SPS1a model point of mSUGRA parameter space and masses were reconstructed 
with the same precision as for the SU3 point.  

The approach developed in this paper can be used for the ATLAS 
and CMS detectors at the LHC; or at the future ILC.

\section*{Acknowledgments}  
The authors thank K.Cranmer, M.Ibe, C.G.Lester,
I.Logashenko, A.Mincer, P.Nemethy, F.Paige, and A.R.Raklev for interesting discussions and useful suggestions.
This work has been supported by the National Science Foundation under grants PHY 0428662, PHY 0514425
and PHY 0629419.

\newpage

\section*{Appendix A. Mass relation equation}

The mass relation method \cite{nojiri} is based on a 
solution of a system
of kinematic equations obtained for each vertex of decay cascade.
For the chain (\ref{chain}) the system of kinematic equations is given by
four equations corresponding to four vertices

$m_{\tilde{l}_{R} }^2 = (p_{\tilde{\chi}_{1}^{0}} + k_{l_{1}})^{2}$

$m_{\tilde{\chi}_{2}^{0} }^2 = (p_{\tilde{\chi}_{1}^{0}} + k_{l_{1}} + k_{l_{2}})^{2}$

$m_{\tilde{b} }^2 = (p_{\tilde{\chi}_{1}^{0}} + k_{l_{1}} + k_{l_{2}} + p_{b_{1}})^{2}$

$m_{\tilde{g} }^2 = (p_{\tilde{\chi}_{1}^{0}} + k_{l_{1}} + k_{l_{2}} + p_{b_{1}} + p_{b_{2}})^{2}$

This system of four equations can be easily modified to a linear system 
defining four component of lightest neutralino four-momentum

\begin{equation}
Sp_{\tilde{\chi}_{1}^{0}} = Q,
\label{kineq} 
\end{equation}

where matrix S is given by 

\begin{equation}
\nonumber
\begin{pmatrix} 
k_{l_{1}}^{0}  & -k_{l_{1}}^{1} & -k_{l_{1}}^{2} & -k_{l_{1}}^{3} \\ 
k_{l_{2}}^{0}  & -k_{l_{2}}^{1} & -k_{l_{2}}^{2} & -k_{l_{2}}^{3} \\ 
p_{b_{1}}^{0}  & -p_{b_{1}}^{1} & -p_{b_{1}}^{2} & -p_{b_{1}}^{3} \\ 
p_{b_{2}}^{0}  & -p_{b_{2}}^{1} & -p_{b_{2}}^{2} & -p_{b_{2}}^{3}  
 
\end{pmatrix}                
\end{equation}

$p_{\tilde{\chi}_{1}^{0}}$ is the column vector of lightest neutralino four-momentum

\begin{equation}
\nonumber
\begin{pmatrix} 
p_{\tilde{\chi}_{1}^{0}}^{0} \\
p_{\tilde{\chi}_{1}^{0}}^{1} \\
p_{\tilde{\chi}_{1}^{0}}^{2} \\
p_{\tilde{\chi}_{1}^{0}}^{3}
 
\end{pmatrix}                
\end{equation}

Q is the column vector of coefficients

\begin{equation}
\nonumber
\begin{pmatrix} 
b \\
c - k_{l_{1}} \cdot k_{l_{2}} \\
d - p_{b_{1}} \cdot (k_{l_{1}} + k_{l_{2}}) \\
e - p_{b_{2}} \cdot (k_{l_{1}} + k_{l_{2}} + p_{b_{1}})
 
\end{pmatrix}                
\end{equation}

where

\centerline {$b = 0.5*(m_{\tilde{l}_{R} }^2 - m_{\tilde{\chi}_{1}^{0} }^2)$}
\centerline {$c = 0.5*(m_{\tilde{\chi}_{2}^{0} }^2 - m_{\tilde{l}_{R} }^2)$}
\centerline {$d = 0.5*(m_{\tilde{b} }^2 - m_{\tilde{\chi}_{2}^{0} }^2)$}
\centerline {$e = 0.5*(m_{\tilde{g} }^2 - m_{\tilde{b} }^2)$} 

Lepton and quark masses are neglected in these equations. 

The solutions of linear equation (\ref{kineq}) is given by standard formulas

\begin{equation}
p_{\tilde{\chi}_{1}^{0}}^{j} = detS_{j}/detS
\label{comp}
\end{equation} 

where the submatrix $S_{j}$ is formed by substituting elements of vector Q
instead of $j^{th}$ column of matrix S. 

By using on-shell mass condition $m_{\tilde{\chi}_{1}^{0}}^{2} = p_{\tilde{\chi}_{1}^{0}}^{2}$
for the lightest neutralino with momentum components from (\ref{comp}) one gets the 
following equation

\begin{equation}
f = (detS_{0}/detS)^{2} -  \sum_{j=1}^{j=3} (detS_{j}/detS)^{2} - m_{\tilde \chi_{1}^{0}}^{2} = 0.
\label{massrel}
\end{equation} 

Equation (\ref{massrel}) is the basic equation of mass relation method and 
it gives the relation between masses of five SUSY particles. Each event is
represented as a curve in five dimensional mass space.
Coefficients of this equation are functions of the  
four-momenta of detected particles where b quarks are measured as jets in the detector. 

\end{document}